Bell's inequalities II: logical loophole in their interpretation

Louis Sica

Code 5630
Naval Research Laboratory
Washington, D. C. 20375
U.S.A.

Ph: 202-767-9466
Fax:202-767-9203
sica@ccf.nrl.navy.mil

Abstract

Assumed data streams from a delayed choice gedanken experiment must satisfy a Bell's identity independently of locality assumptions. The violation of Bell's inequality by assumed correlations of identical form among these data streams implies that they cannot all result from statistically equivalent variables of a homogeneous process. This is consistent with both the requirements of arithmetic and distinctions between commuting and noncommuting observables in quantum mechanics. Neglect of these distinctions implies a logical loophole in the conventional interpretation of Bell's inequalities.

The companion paper Part I [1] (a preliminary version of material in this paper was presented in Ref. [2]) presents a derivation of Bell's inequality [3-10] in which it is shown to be dependent on purely arithmetic assumptions plus the concept of limits. It is here suggested that its derivation from such simple assumptions provides an explanation of the fact that the same result has been derived from a number of different suppositions in the literature. The physical and statistical assumptions that have been folded into previous derivations at various points are largely extraneous, although this is not apparent until the same result is obtained without them.

Bell's inequalities, Part I [1] considered the question of how experiments can violate Bell's inequalities, and showed that for the finite data that characterizes all experiments, Bell's inequalities is an identity in both the three correlation and four correlation cases. These identities may be violated if is not noticed that data for two correlations determines the third in the three correlation case, and that data for three correlations determines the fourth, in the four correlation case. The computation of these correlations from over-determined data sets in multiple runs without data matching (as required by all derivations known to the author) breaks the link between initial assumptions and their derived logical consequences, and thus leads to possible violation of the inequalities. (If a statement violates the assumptions of a derivation, it may not be concluded thereby, that it must either agree or disagree with the conclusions.)

Part I [1] treats requirements for consistency with the derivation where all correlations are computed from data independently of theoretical models for the correlations. By contrast, the present paper considers Bell's gedanken experiment in which the correlation $<ab>$ is measured or assumed known, and the other correlations are constructed from theoretical reasoning. This involves more subtle arguments than those given in Part I [1], but these must be considered in the context of the widely held belief that quantum mechanics violates Bell's inequalities. The situation is logically different in the three correlation and four correlation cases, and they will be treated separately. This is because the issue of nonlocality plays a different role in the two cases.

In this paper, Bell's identity and Bell's inequality will be applied first to a delayed choice gedanken experiment for which three variables are appropriate, whether or not instantaneous nonlocal influences are assumed, and which would therefore yield data satisfying Bell's identity. Bell's inequality, must then be satisfied by theoretically derived correlations among the variables as a consistency condition with arithmetic. The violation of this consistency condition leads to the identification of a widely used but unexamined hidden assumption in the usual application of Bell's inequalities: that correlations among the observables may be considered individually, and all have the same cosine of angular difference form. However, in Part I [1] it has been shown already that one of the correlations is arithmetically dependent on the data of the others, thus violating the basis for the assumption. The present paper extends this result by a consideration of the gedanken experiment interpretation of Bell's inequalities. In this case, commutation of some observables and noncommutation of others (the quantum viewpoint) and the conditional dependence of the unperformed measurements on those actually performed (the hidden variables view) both undermine the equivalence assumption. This issue has rarely been considered (this was noted, however, in Ref. [11]) in the interpretation of Bell's inequalities, and its omission amounts to an important logical loophole in the conventional treatment.

The basic facts derived in Part I [1] for Bell's original equality are the following. Assume that there exist three lists of numbers, each of length $N$, with each number restricted to the values $\pm 1$. The lists are denoted $a$, $b$, and $b'$ and their respective members by $a_i$, $b_i$, and $b_i'$, $i = 1...N$. Then Bell's identity holds:

$$\left| \sum_{i=1}^{N} a_i b_i / N - \sum_{i=1}^{N} a_i b_i' / N \right| \leq 1 - \sum_{i=1}^{N} b_i b_i' / N \quad . \tag{1}$$

Since the truth of (1) is independent of $N$, the limits $N \to \infty$ may be taken, assuming that they exist in some sense for the numerical sequences used, and each numerical average may be replaced by a corresponding ensemble average. If the standard bracket notation for ensemble average is used, one arrives at the original form for Bell's inequality [3]:

$$|<ab> - <ab'>| \leq 1 - <bb'> \quad . \tag{2}$$

Inequality (2), must be carefully distinguished from precursor identity (1) as to both meaning and use. While identity (1) cannot be violated by any finite sets of $\pm 1$'s, whether random or non random, (2) is not an identity in formulas that may be inserted into it, and hence *may* be violated by derived but theoretically inconsistent formulas for limits of averages. Of course, if the ensemble averages in (2) are literally the limits of the averages in (1), then they must automatically satisfy (2). Inequality (2) thus provides a constraint that functions representing theoretically computed ensemble averages must satisfy for internal consistency with any data sets that could possibly exist or be imagined. This is entirely apart from the question of whether they accurately describe the results of experiments. These relationships seem not to have been stated previously and are central to the analysis that follows.

In applying (2) to observations, three variables must be logically identified from a physical situation. In the measurement of spin components of two spin-1/2 particles in a singlet state flying apart (Fig. 1), the apparatus is assumed to be run in a delayed choice mode with angular settings of Stern-Gerlach magnets made on the fly. It is assumed here that the measurement at *a* is completed before that at *b* or *b'* has begun, so that causal effects resulting from apparatus settings can only travel from *a* to *b* or *b'* but not from *b* to *a*. The *a* and *b* measuring devices are separated by a distance large compared to that which light can traverse in the time between measurements.

For a given orientation of the magnet *a*, the magnet *b* may also be imagined to be in a second, different orientation, *b'*, thus resulting in a different stream of data for the same sequence of hidden variables used to determine *b* (and *a*). This data-stream is unacquired however, since the magnet producing the field orientation corresponding to *b*, being a classical macroscopic object, can only have one angular position at a time. As a result, no actual experiment can be performed with this apparatus that tests the complete inequality (2), since only one pair of values for *a* and *b* can be obtained per particle pair. The outputs of variables *a*, *b*, and *b'* can be measured in neither a single experiment (this would not be possible classically either), nor in a sequence of experiments as would be possible classically if the initial conditions and assumed hidden variables were known and controlled.

In Bell's prescription [3], *a* and *b* are measured allowing the correlation <*ab*> to be computed, and the other correlations are inferred from theoretical reasoning based on the assumption that they are functions of hidden variables. (Bell's definition of correlation is the negative of that used here. Bell assumed, in fact, that all variables were based on the same hidden variable readout function.) Bell's insight resides in the fact that while the variables *a*, *b*, and *b'* are not all measurable from one particle pair using the (classical) experimental apparatus in Fig.1, their values could all be predicted classically for two successive experiments as a function of fixed initial conditions, hidden variables, and different apparatus settings even though infinite precision might be necessary in a chaotic situation. Thus, all the correlations could be computed for this apparatus, for successive classical experiments and appropriate ensemble averaging, even though they do not exist at the same time for a single apparatus setting.

Under the conditions of the delayed choice gedanken experiment described, readout values at *b* and *b'* would be affected if a nonlocal influence on them due to *a* were assumed, but three data streams would still result. Three data streams would also result if no nonlocal influence were assumed. Thus, in either case, Bell's identity for three experimental correlations would be satisfied by the three finite length data streams. This result, somewhat different from the four correlation case, is the motivation for considering the three correlation case.

Thus, once hidden variables in the above sense are assumed to exist, the only question is what form the correlations in (2) take. It is well known experimentally and from quantum mechanics, that for the two data streams *a* and *b* produced by detector orientations (*a*) and (*b*), <*ab*> = - cos[ (*b*) - (*a*)]. Owing apparently to the symmetry of the apparatus, the additional correlations <*ab'*> and <*bb'*> of unknown and un-acquired readouts have been widely assumed to be given by the same cosine of angular differences equation, i.e. the same function as <*ab*>. This assignment of correlation functions is assumed to result from quantum mechanics. For certain values of the three angles, (2) is then violated.

The failure of the assigned correlations to satisfy (2) indicates either that the assumed alternate data cannot exist, or if it does exist, that the assumption that all variables and their correlations are equivalent is flawed. It is important to realize that it is not readout functions per se, in the sense of Bell, that are prohibited, since any literal values whatsoever for *b'* may be used in conjunction with real data for *a* and *b* having a cosinusoidal correlation without violating identity (1), and if the limits exist, inequality (2). However, the other two correlations cannot then have the same cosine form as <*ab*>, and for the present delayed choice experiment this follows whether or not measurements of *a* are assumed to influence *b* and *b'* or not. Thus, correlations among the performed and unperformed measurements cannot all have the same cosine form if hidden variables exist at all.

However, in addition to violating the mathematical consistency condition (2), the assumption of equivalent variables violates quantum mechanics, since the quantum mechanical operators corresponding to spin measurements at *b* and *b'* for the same particle don't commute, while those at *a* and *b*, measurements on *two different particles*, do commute. The real measurements are thus on a different statistical footing from the imaginary ones, and the correlations among the variables surely cannot all be assumed to be given by the same function in the absence of proof that this is the case. Noncommuting observables occupy an entirely different place in conventional quantum mechanics from commuting observables (see virtually any quantum mechanics text), and hidden variables if assumed to exist, must duplicate such statistical properties as quantum mechanics specifies (note also the related fact of classical mechanics that two perpendicular components of angular momentum cannot be simultaneous canonical momenta) [12].

Ultimately, due to the fact of noncommutation, the correlations required for insertion in inequality (2) are not really computable from quantum mechanics, and require a specific hidden variables theory for evaluation. Bell indicated [13] that $b'$ was to be considered as a possible alternative setting of $b$ under the condition of identical hidden variable values. (This is equivalent to performing the experiment again under identical conditions and with alternate setting $b'$.) But once a specific value for $b$ is obtained, the range of values of any hidden variables leading to that value is partially limited, so that any values at alternative setting $b'$ are conditionally dependent on it. Thus from the hidden variables viewpoint, the correlations should not be considered to be equivalent.

It seems to be believed that non-commutation is a property of the quantum world only. However, simple reflection indicates that this belief is unjustified, and one can find numerous instances in which alternative orderings of classical operations produce different results. A well known example is that finite rotations of solid objects in three dimensions do not commute. In addition, the implications of non commuting operations for stochastic processes have not been explored and made a part of classical physics as they have quantum mechanics.

The reasoning for the four correlation inequality is similar to that for three correlations but with some differences. It was shown in Part I [1] that for four streams of data of length $N$ restricted to $\pm 1$, Bell's identity is

$$\left| \frac{1}{N}\sum_{i=1}^{N} a_i b_i + \frac{1}{N}\sum_{i=1}^{N} a_i b_i' \right| + \left| \frac{1}{N}\sum_{i=1}^{N} a_i' b_i - \frac{1}{N}\sum_{i=1}^{N} a_i' b_i' \right| \leq 2 \qquad (3)$$

and Bell's inequality is:

$$|\langle ab\rangle + \langle ab'\rangle| + |\langle a'b\rangle - \langle a'b'\rangle| \leq 2 \tag{4}$$

Comments following (2) above apply to (4) as well. It is not an identity, but a consistency condition on the functions of which it is composed in order that they represent correlations among any four (infinite) lists of ± 1's that could possibly exist.

If a delayed choice gedanken experiment is considered for two detector positions on the A side and two on the B side (See Fig.1), four data streams result if locality is assumed, and six if nonlocal effects are assumed. The later assumption is thus inconsistent with the use of Bell's identity for four correlations. This assumption is not logically required, however. It was considered in the three correlation case to illustrate in a graphic way that the assumption of the statistical equivalence of variables is sufficient in and of itself to cause a Bell's inequality violation independently of whether or not the nonlocality assumption is made. In the four variable case which does require the assumption of locality to obtain four data streams, violation of Bell's inequality by the correlations then depends on the correlation equivalence assumption. If that assumption is flawed, violation of Bell's inequality no longer has the same paradoxical implications.

It has been shown in Part I [1] that the correlation equivalence assumption is numerically unjustified in the four correlation case, since data for three correlations determine the fourth. Related reasoning applies to the gedanken experiment also. Definite values of *a* and *b* constrain the values of assumed underlying hidden random variables that in turn determine *a'* and *b'*. Consequently, as discussed in the three correlation case, the probability for obtaining particular values of *a'* and *b'* must be conditional on the values of *a* and *b*, and the forms of the correlations may thereby be affected.

The reasoning from quantum mechanics is different but related, and yields conclusions harmonious with the above. While *a* and *b* commute, *a* and *a'* do not, and *b* and *b'* do not. The statistics of hidden variable models must duplicate those of quantum mechanics by definition. Thus, reasoning based on the requirements of quantum mechanics, probability considerations, and arithmetic is consistent in indicating that the correlations of the four variables should not be assumed to be functionally the same. This indicates a reason for Bell's inequality violation that has been largely unconsidered previously. As indicated, however, a specific hidden variables theory is necessary to evaluate all the correlations used in the four correlation inequality in the sense of Bell. The conclusions drawn in the present work are consistent with that of Ref. [14], i.e. that a joint probability density function for all variables in the inequality must exist for Bell's inequality to be valid.

The assumption that all correlations in Bell's inequalities are equivalent, reduces the generality of hidden variables models that may be considered. This may be seen by using the conventional terminology of stochastic processes to express the implications of Bell's inequality violation. Consider a process such as a random rectangular wave with values equal to +1 or − 1 as shown in Fig. 2, but with otherwise unspecified statistical properties. The schematic waveform shown is a member of an ensemble of waveforms with parameters controlled by some unspecified probability distribution. Suppose that the process is homogeneous in the spatial coordinate $x$, i.e. wide sense stationary in that coordinate. Then by definition, values of the waveform may be read out simultaneously at locations $x_1$, $x_2$, and $x_3$ for each of $N$ realizations, and the correlations $C(x_3-x_2)$, $C(x_2-x_1)$, and $C(x_3-x_1)$ may then be calculated from these data sets for the ensemble of realizations as $N \rightarrow \infty$. The correlations so computed are given by the same function $C(x)$ evaluated at different values of its argument, and this function may also be evaluated using a separate run for each value of its argument. The data streams at locations $x_1$, $x_2$, and $x_3$ for $N$ realizations of the process must satisfy Bell's identity (1), and their correlations must satisfy Bell's inequality (2). However, if the correlations are cosines, then Bell's inequality (2) will be violated. Thus, it follows that a process quantized to ±1 with correlation given by the cosine of coordinate differences cannot be represented by a homogeneous process. However, homogeneous processes represent a subset of those classified in books on random processes [15,16].

The lack of examination of the assumption of statistical equivalence of the correlations among the variables amounts to a previously unidentified logical loophole in the interpretations of Bells' inequalities. The analysis of the three correlation case, in particular, lends support to the conclusion that nonlocality is not necessarily implicated as the reason for violation of Bell's inequalities. The result in the four correlation case is not as strong, since the assumption of locality is now required, but once the assumption of statistical equivalence of correlations is removed, no paradoxical violation of the inequalities necessarily follows. Further, the example of Part I [1], Fig. 2 shows the case of an experiment for which four variables may be explicitly measured for a single particle pair, leading to satisfaction of Bell's inequalities. The computed correlations do not all have the same functional form and are not in general zero.

Ultimately, only a subset of random variables models for quantum correlations appear to be excluded by Bell's inequalities violation. These results leave open the question of whether or not the negative cosine correlation of spin measurements for particles in the singlet state implies an intrinsically nonlocal model for reasons other than those tied to Bell's inequality violation. However, Bell knew, in spite of the example of Bohmian mechanics, that his failure to find a local model for this correlation was not proof that one could not exist. It was in an attempt to prove such non-existence that he constructed his inequalities. It is worth noting, in this regard that a model has recently been demonstrated by Steiner [17] in which the necessary nonlocal information transmitted to simulate the cosine correlation is only 1.48 bits. No lower limit has been proven in conjunction with this demonstration.

**Acknowledgements**

I gratefully acknowledge Michael Steiner for many stimulating conversations on the subject of this article and important critical comments on the manuscript. I am also indebted to Yanhua Shih and Carlton M. Caves for critical comments on an early version of the manuscript, and to Kent S. Wood for useful suggestions regarding presentation of the material. Finally, I would like to thank Herschel S. Pilloff for continued help and encouragement during the course of this study

Figure Captions

1. Schematic Stern-Gerlach apparatus. Arrows a, b, indicate magnetic field directions encountered by pairs of particles emitted in opposite directions by the source. Distances from source to measuring devices are shown unequal to allow a-measurement to be performed first. At each encounter with a magnetic field, the particle is deflected in one of two directions depending on whether its spin is +1 or -1 (half). The direction b' indicates an alternative direction of b for which data would result if that direction were selected. Similarly, an alternative a' direction (not shown) could be considered.

2. Single realization of a hypothetical random, quantized, waveform from a homogeneous process sampled at $x_1$, $x_2$, and $x_3$ for each realization.

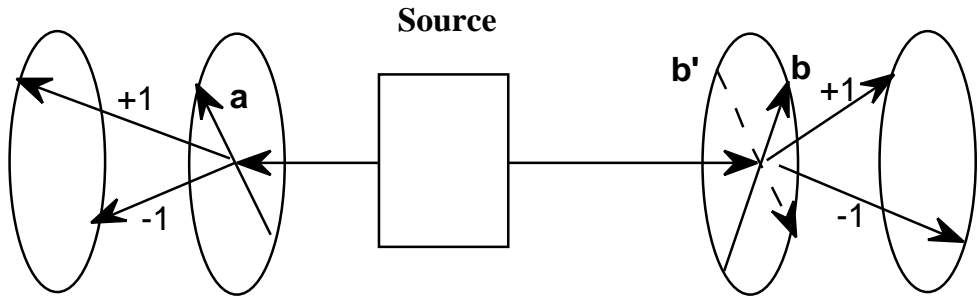

Fig. 1

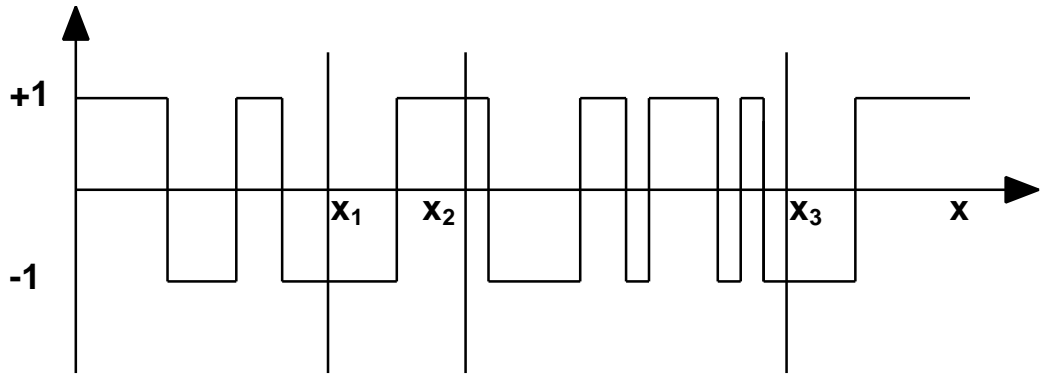

Fig 2